# Borophene hydride: a stiff 2D material with high thermal conductivity and attractive optical and electronic properties

Bohayra Mortazavi[*,1], Meysam Makaremi[2], Masoud Shahrokhi[3], Mostafa Raeisi[4], Chandra Veer Singh[2,5], Timon Rabczuk[1,6], Luiz Felipe C. Pereira[#,7]

[1]*Institute of Structural Mechanics, Bauhaus-Universität Weimar, Marienstr. 15, D-99423 Weimar, Germany.*
[2]*Department of Materials Science and Engineering, University of Toronto, 184 College Street, Suite 140, Toronto, ON M5S 3E4, Canada.*
[3]*Institute of Chemical Research of Catalonia, ICIQ, The Barcelona Institute of Science and Technology, Av. Països Catalans 16, ES-43007 Tarragona, Spain.*
[4]*Mechanical Engineering Department, Imam Khomeini International University, PO Box: 34149-16818, Qazvin, Iran*
[5]*Department of Mechanical and Industrial Engineering, University of Toronto, 5 King's College Road, Toronto M5S 3G8, Canada*
[6]*College of Civil Engineering, Department of Geotechnical Engineering, Tongji University, Shanghai, China.*
[7]*Departamento de Fisica, Universidade Federal do Rio Grande do Norte, 59078-970 Natal, Brazil*

**Abstract**

Two-dimensional (2D) structures of boron atoms so called borophene, have recently attracted remarkable attention. In a latest exciting experimental study, a hydrogenated borophene structure was realized. Motivated by this success, we conducted extensive first-principles calculations to explore the mechanical, thermal conduction, electronic and optical responses of borophene hydride. The mechanical response of borophene hydride was found to be anisotropic in which it can yield an elastic modulus of 131 N/m and a high tensile strength of 19.9 N/m along the armchair direction. Notably, it was shown that by applying mechanical loading the metallic electronic character of borophene hydride can be altered to direct band-gap semiconducting, very appealing for the application in nanoelectronics. The absorption edge of the imaginary part of the dielectric function was found to occur in the visible range of light for parallel polarization. Finally, it was estimated that this novel 2D structure at the room temperature can exhibit high thermal conductivities of 335 W/mK and 293 W/mK along zigzag and armchair directions, respectively. Our study confirms that borophene hydride benefits an outstanding combination of interesting mechanical, electronic, optical and thermal conduction properties, promising for the design of novel nanodevices.

Corresponding authors: *bohayra.mortazavi@gmail.com, ltimon.rabczuk@uni-weimar.de, #pereira@fisica.ufrn.br

**1. Introduction**

Outstanding properties of graphene [1,2], motivated the synthesis of other two-dimensional (2D) materials to discover new physics, with a goal to fabricate advanced materials with enhanced performance. During the last decade, the family of 2D materials has been growing continuously which includes prominent members like hexagonal boron-nitride [3,4], silicene [5,6], germanene [7], stanene [8], transition metal dichalcogenides [9–11], carbon-nitride 2D structures like g-$C_3N_4$ [12,13], $C_2N$ [14], $C_3N$ [15] and phosphorene [16,17].

Despite the great experimental success in the fabrication of graphene-like 2D materials which include a rich, diverse and growing family, only a few 2D materials are monoelemental, such as phosphorene, silicene, germanene and stanene. Notably these monoelemental 2D materials show out-of plane buckled structures, different from graphene with its stable flat form. The main reason behind such a structural difference is that these elements present limited ability in varying their bonding nature unlike carbon with its versatile bonding such as $sp^3$, $sp^2$ and sp [18]. Nevertheless, boron, the neighboring element of carbon placed between metals and non-metals in the periodic table, can also present diverse bonding and as such it is available in a variety of structures from zero-dimensional to three-dimensional [18–20]. Interestingly, during the last couple of years, three different 2D boron structures, with the both flat [21] and buckled [22] geometries have been successfully synthesized through molecular beam epitaxy growth on a silver surface under ultrahigh-vacuum conditions. These recent experimental advances in the fabrication of borophene stimulates theoretical studies to explore its applications in various systems such as hydrogen storage, rechargeable metal-ion batteries, superconductors and mechanically robust components [23–44].

The research on the 2D boron structures is yet at very early stages and there are numerous experimental challenges. Among these, the greatest obstacle is to develop efficient transferring methods for lifting the borophene sheets from the silver substrate in order to reach isolated nanomembranes. Most recently, experimental realization of 2D hydrogen boride sheets with an empirical formula of $B_1H_1$ was successfully achieved by exfoliation and complete ion-exchange between protons and magnesium cations in magnesium diboride [45]. The fabricated hydrogenated borophene structures were in multi-layer form and remarkably were not grown on a substrate which can accordingly serve as a promising sign toward their practical applications. It is worthy to note that the existence and stability of such a 2D structure was already theoretically predicted [46]. This exciting latest experimental advance [45], raises the importance of intensive theoretical studies to establish a comprehensive knowledge about different material properties, very critical for the design of nanodevices. To the best of our knowledge, the mechanical/failure and thermal conduction properties of borophene hydride 2D material have not been studied. Moreover, the effects of different loading conditions on the evolution of electronic and optical properties of this novel 2D structure have not been also investigated. Accordingly, in the present investigation our objective is to explore the mechanical, electronic, optical and thermal conductivity, properties of this recently synthesized hydrogenated borophene, via extensive first-principles density functional theory (DFT) calculations.

**2. Computational details**

The DFT calculations in this study were performed using the Vienna ab-initio simulation package (VASP) [47–49]. A plane wave basis set with an energy cut-off of 500 eV and the generalized gradient approximation exchange-correlation functional proposed by Perdew-Burke-Ernzerhof [50] were employed. VMD [51] and VESTA [52] packages were also used for the visualization of atomic structures. Periodic boundary conditions were applied along all three Cartesian directions and in order to avoid image-image interactions along the sheets normal direction a vacuum layer of 20 Å was employed. To obtain the energy minimized structure, a unit-cell with 4 boron and 4 hydrogen atoms was considered. The energy minimized structure was obtained by changing the size of the unit-cell and then employing the conjugate gradient method for the geometry optimizations, with termination criteria of $10^{-6}$ eV and 0.005 eV/Å for the energy and forces, respectively. We also used 19×19×1 Monkhorst-Pack [53] k-point mesh size. Similarly to graphene, $B_1H_1$ also shows two major orientations, which one can refer to armchair and zigzag directions in analogy to graphene. We therefore particularly analyzed anisotropy in the mechanical, optical and thermal conductivity responses of single-layer borophene hydride.

Since the dynamical effects such as the temperature are not taken into consideration and periodic boundary conditions were also applied along the planar directions, only a unit-cell modelling [54] is accurate enough for the evaluation of mechanical properties, and we therefore used a unit-cell consisting of 8 atoms. To evaluate the mechanical properties, we increased the periodic simulation box size along the loading direction in a multistep procedure, every step with a small engineering strain of 0.0005. For the uniaxial loading conditions, upon the stretching along the loading direction the stress along the transverse direction should be negligible. To satisfy this condition, after applying the loading strain, the simulation box size along the transverse direction of loading was changed accordingly in a way that the transverse stress remained negligible in comparison with that along the loading direction. For the biaxial loading condition, the equal loading strain was applied simultaneously along the both planar directions. After applying the changes in the simulation box size, atomic positions were rescaled to avoid any sudden structural changes. We then used the conjugate gradient method for the geometry optimizations, with termination criteria of $10^{-5}$ eV and 0.005 eV/Å for the energy and the forces, respectively, in which we used a 15×15×1 k-point mesh size. Final stress values after the termination of energy minimization process were calculated to obtain stress-strain relations.

The ground state electronic properties were first calculated using the PBE functional. After the calculation of electronic ground states, optical properties such as the imaginary and real parts of the dielectric function were acquired by solving the random phase approximation (RPA) [55] plus PBE. In the RPA approach, electrons are assumed to respond only to the total electric potential which is the sum of the external perturbing potential and a screening potential. The external perturbing potential is assumed to oscillate at a single frequency (ω), so through a self-consistent field (SCF) [56] method. This model yields a dynamic dielectric function denoted by ε(ω). Optical properties are determined by the dielectric function $\varepsilon(\omega) = \mathrm{Re}\,\varepsilon_{\alpha\beta}(\omega) + i\,\mathrm{Im}\,\varepsilon_{\alpha\beta}(\omega)$, which is dependent on the electronic structure, and can be obtained from following relations [57,58]:

$$\mathrm{Im}\,\varepsilon_{\alpha\beta}(\omega) = \frac{4\pi^2 e^2}{\Omega} \lim_{q \to 0} \frac{1}{|q|^2} \sum_{c,v,k} 2w_k \delta(\varepsilon_{ck} - \varepsilon_{vk} - \omega) \times \langle u_{ck+e_\alpha q} | u_{vk} \rangle \langle u_{ck+e_\beta q} | u_{vk} \rangle^* \quad (1)$$

where $q$ is the Bloch vector of the incident wave and $w_k$ the **k**-point weight. The band indices $c$ and $v$ are restricted to the conduction and the valence band states, respectively. The vectors $e_\alpha$ are the unit vectors for three Cartesian directions and $\Omega$ is the volume of the unit cell. $u_{ck}$ is the cell periodic part of the orbitals at the $k$-point **k**. The real part $\mathrm{Re}\,\varepsilon_{\alpha\beta}(\omega)$ can be evaluated from $\mathrm{Im}\,\varepsilon_{\alpha\beta}(\omega)$ using the Kramers–Kronig transformation:

$$\mathrm{Re}\,\varepsilon_{\alpha\beta}(\omega) = 1 + \frac{2}{\pi} P \int_0^\infty \frac{\omega' \mathrm{Im}\,\varepsilon_{\alpha\beta}(\omega')}{(\omega')^2 - \omega^2 + i\eta} d\omega' \quad (2)$$

Local field effects, which correspond to changes in the cell periodic part of the potential, were included in the random phase approximation.

The lattice thermal conductivity of single-layer $B_1H_1$ borophene hydride was studied using the ShengBTE [59] package, which implements a fully iterative solution of the Boltzmann transport equation. Second order (harmonic) and third-order (anharmonic) interatomic force constants were calculated from first principles calculations based on the DFT. A fully iterative solution, such as the one implemented in ShengBTE is expected to yield more accurate results than the solutions based on the single mode relaxation time approximation (SMRTA). Phonon frequencies and the harmonic interatomic force constants were obtained based on the density functional perturbation theory (DFPT) results using the PHONOPY code [60] for a 3×5×1 super-cell with 5×5×1 k-point grid. The third-order anharmonic force constants were calculated using the finite displacement approach for the same super-cell and k-point mesh size. For the anharmonic force constants, we include the interactions with the tenth nearest-neighbour atoms. The convergence of the thermal conductivity with respect to $q$-points in the full Brillouin zone sampling was confirmed and in order to report the final thermal conduction properties we used a 50×50×1 $q$-mesh. We also calculated the Born effective charges and dielectric constants using the DFPT method and they were considered in the dynamical matrix as a correction to take into account the long-range electrostatic interactions.

## 3. Results and discussions

The atomic structure of energy minimized borophene hydride is illustrated in Fig. 1, which shows a graphene-like honeycomb lattice made from boron atoms functionalized by the hydrogen atoms. In this structure, the layer of pure boron atoms is sandwiched between two layers of hydrogen atoms, which are decorated on the both sides of the B-B bonds. The unit-cell lattice constants were acquired to be 5.290 Å and 3.018 Å which match closely with the values of 5.299 Å and 2.988 Å reported in the original theoretical work by Jiao *et al.* [46]. In Fig. 1, we also plotted the electron localization function (ELF) [61] which takes a value between 0 and 1, where ELF=1 corresponds to perfect localization. In the lattice of $B_1H_1$, two different B-B bonds exist. For the B-B bonds without the hydrogen bridging, the electron localization occurs at the center of B−B bonds which is the characteristic of covalent bonding (Fig. 1 top view). However, for the other B-B bonds bridged by the hydrogen atoms the electrons are localized around the H atoms (Fig. 1 side view). For the energy minimized structure, the bond length of B-B bonds without and with hydrogen bridging were measured to be 1.720 Å and 1.820 Å, respectively, and the B-H bond length was found to be 1.328 Å.

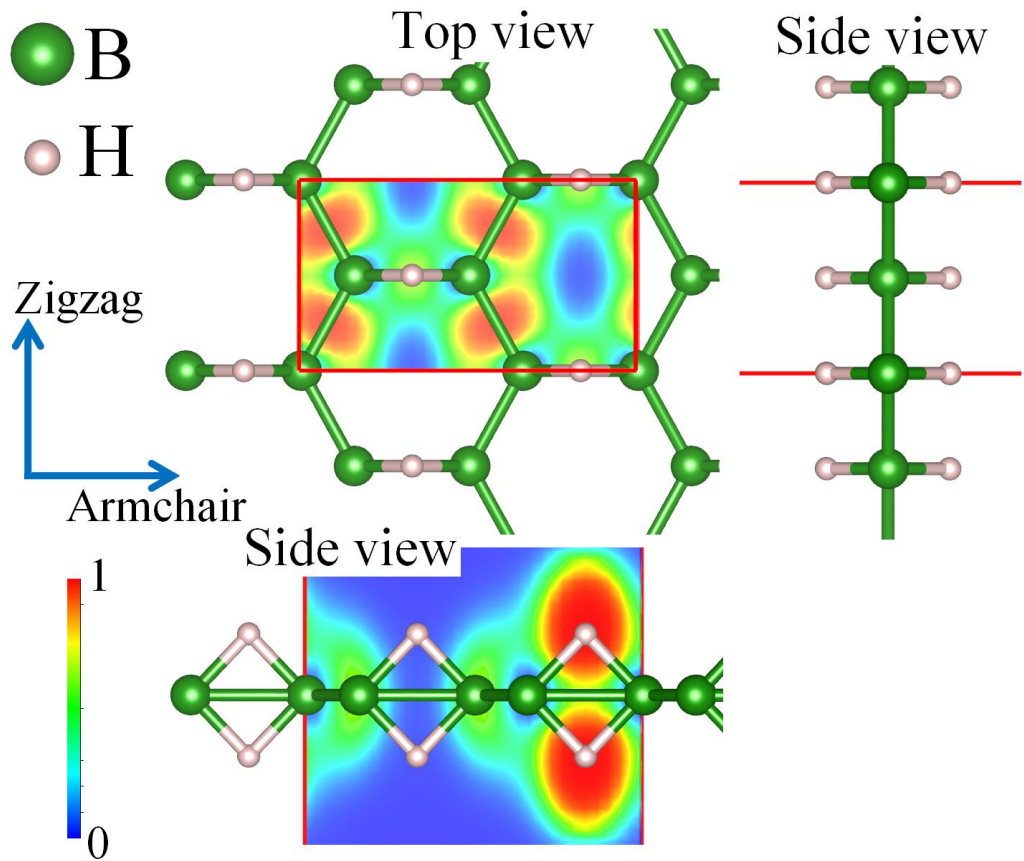

Fig. 1, Top and side views of atomic configuration in single-layer $B_1H_1$. The box with the red colour shows the $B_4H_4$ unit-cell which was used to evaluate mechanical, electronic and optical properties. We studied the properties along the armchair and zigzag directions as shown. The contours illustrate electron localization function (ELF) plotted on the two sections.

In Fig. 2a, DFT predictions for the biaxial and uniaxial stress-strain responses of borophene hydride, elongated along the armchair and zigzag directions are plotted. In all cases, stress-strain responses present an initial linear relation which is followed by a nonlinear trend up to the ultimate tensile strength point, a point at which the material illustrates its maximum load bearing. The slope of the first initial linear section of the uniaxial stress-strain response is equal to the elastic modulus. In this work we therefore fitted a line to the uniaxial stress-strain values for the strain levels below 0.02 to report the elastic modulus. Based on our modelling results, the elastic modulus of borophene hydride along the armchair and zigzag directions were calculated to be 131 N/m and 99 N/m, respectively. For these initial strain levels within the elastic limit, the strain along the traverse direction ($s_t$) with respect to the loading strain ($s_l$) is acceptably constant and can be used to obtain the Poisson's ratio using the $-s_t/s_l$. This way Poisson's ratio along the armchair and zigzag directions were predicted to be 0.25 and 0.19, respectively. The tensile strength along the armchair and zigzag directions were found to be 19.9 N/m and 17.8 N/m, respectively. Our results shown in Fig. 2a confirms that the both linear and non-linear parts of uniaxial stress-strain curves are distinctly different which suggests anisotropic mechanical response of $B_1H_1$ nanosheets in which along the armchair and zigzag the structure shows higher rigidity and stretchability, respectively.

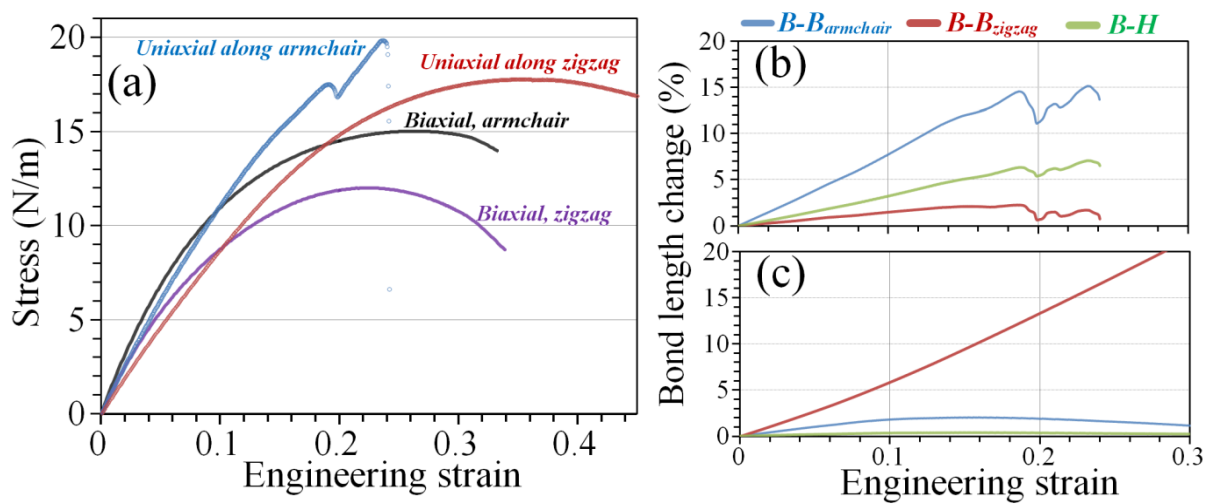

Fig. 2, (a) Uniaxial and biaxial stress-engineering strain responses of single-layer and free-standing borophene hydride. The change in the bond lengths for the uniaxial loading along the (b) armchair and (c) zigzag direction.

As it is clear, the higher elastic modulus and tensile strength along the armchair direction can be partly attributed to the stiffening effect of B-H bonds that are exactly oriented along the armchair direction. To better understand the underlying mechanism that results in anisotropic tensile response of borophene hydride, we analyzed the evolution of bond lengths. Fig. 2b and Fig 2c, compares changes in the bond lengths as a function of strain for the uniaxial loading along the armchair and zigzag direction, respectively. In this system, there are three different bonds, B-B covalent bonds (oriented along zigzag), B-B bonds along armchair (which includes hydrogen bridging) and B-H bonds. For the stretching along the armchair direction, two bonds are exactly along the loading direction and directly involve in the load bearing and such that by increasing the engineering strain level these bonds increases substantially which results in higher stress values and elasticity. In this case, the bond oriented along the transverse direction of loading also increases. On the other side, for the uniaxial stretching along the zigzag, only the covalent B-B bonds are almost oriented along the loading direction and the other bonds in the system are exactly along the transverse direction of loading. In this case, according to the results illustrated in Fig. 2c, by increasing the strain level the B-B covalent bond keeps increasing and at the same time the other B-B bond first only slightly stretches and at higher strains of ~0.15 starts to contract. In this case, notably the B-H bond length changes remain below ~2% which further confirms that this bond plays no major role in the load bearing along the zigzag direction. Interestingly, the covalent B-B bonds show high stretchability as compared with the other bonds in the $B_1H_1$ structure which allows the structure to endure at higher strain levels when stretching along the zigzag direction.

The result shown in Fig. 2a for the uniaxial strain along the armchair direction reveals that the structure exhibits an unusual yield point at strain levels around 0.19 in which a slight drop in the stress values are observable. By further increasing the strain levels the stress values start to increase again to reach the maximum tensile strength point at strain levels around 0.24. Our bond analysis depicted in Fig. 2b clearly confirms that this yield point in the stress-strain response is basically due to the bond contractions. If one compares the bond lengths at this initial yield point and tensile strength point, very close values are observable. Our structural analysis reveal that at this yield point the structure contracts considerably along the sheet transverse direction which helps the material to flow easier along the loading direction and such that it relieves the bond stresses by bond contraction. In Fig. 2a, the stress values along the armchair and zigzag directions for the biaxial loading are also plotted which similarly show higher stress values along the armchair direction.

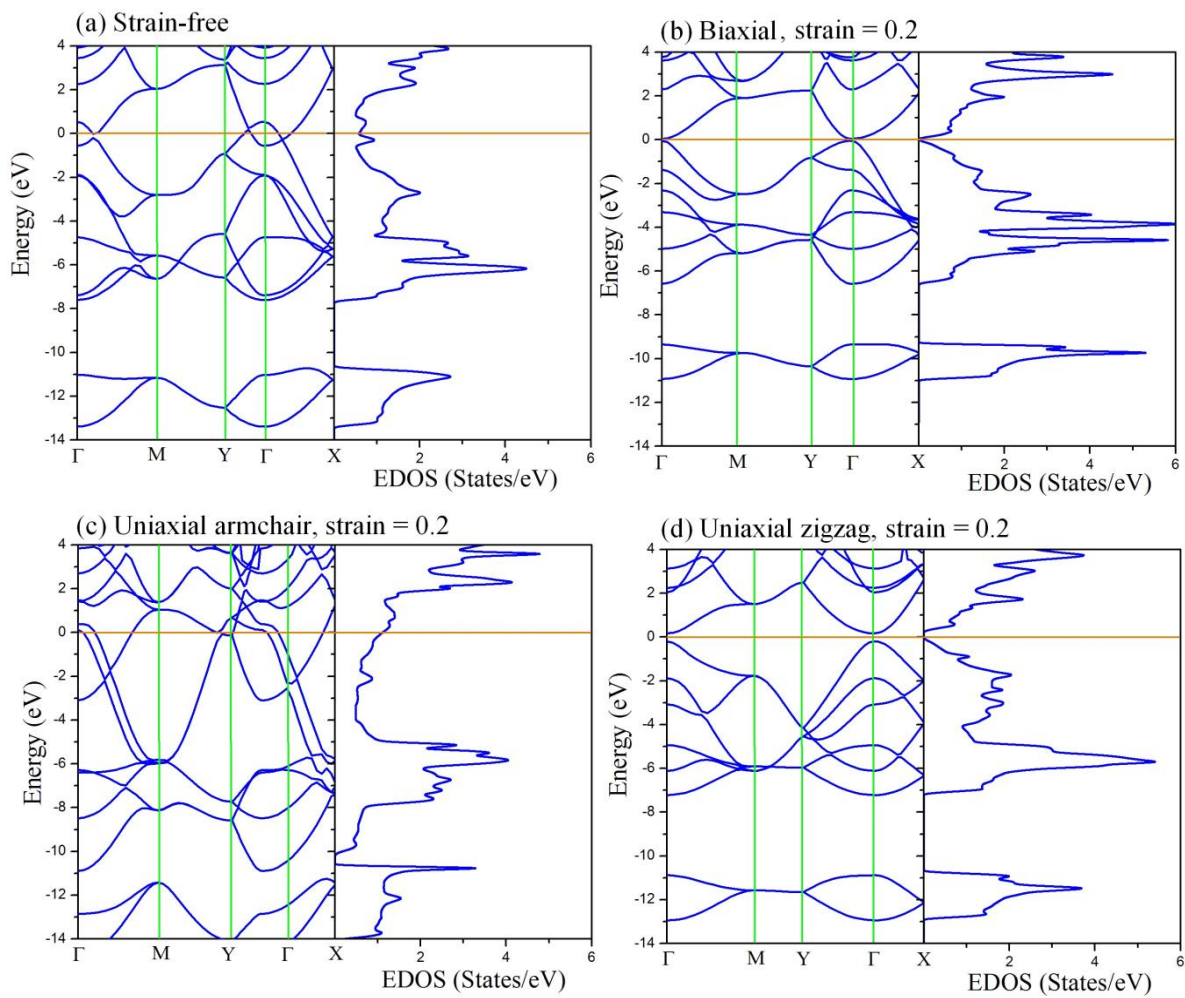

Fig. 3, Band structure and total EDOS of unstrained and strained single-layer borophene hydride predicted by the PBE functional. For the strained structures, the applied engineering strain is equal to 0.2. The Fermi energy is aligned to zero.

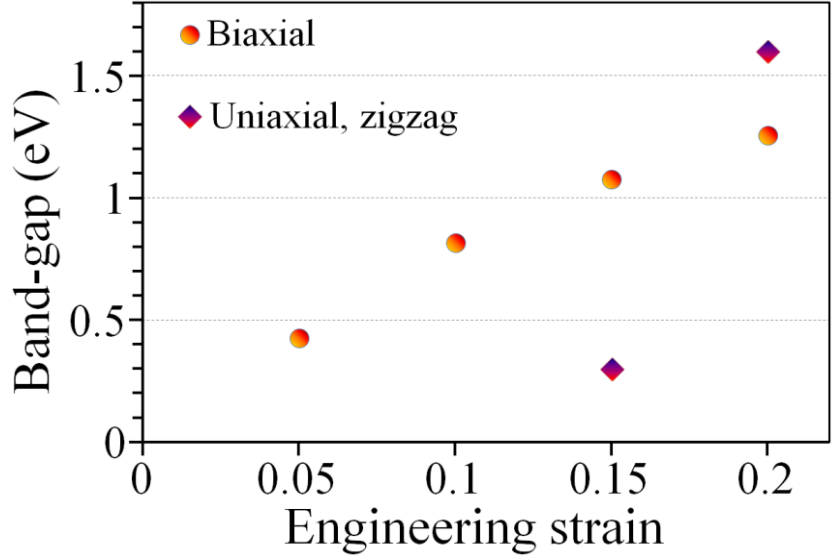

Fig. 4, Electronic band-gap of single-layer borophene hydride as a function of applied engineering strain as predicted by the HSE06 functional.

Next we shift our attention to investigate the electronic and optical properties of monolayer borophene hydride. In this case, we particularly analyzed the possibility of tuning of the electronic and optical properties by uniaxial or biaxial loading. First, in order to probe the electronic properties, the band structure along the high symmetry directions and total electronic density of states (EDOS) are calculated using the PBE functional. Fig. 3a illustrates the band structure and EDOS of strain-free borophene hydride. As shown in this figure, this structure yields metallic character and moreover our predicted band structure match excellently with the original theoretical work [46]. Figs. 3b-d, respectively, illustrate the evolution of band structures and corresponding EDOS of borophene hydride under biaxial loading and uniaxial loading along the armchair and zigzag directions, with an engineering tensile strain of 0.2. Based on our results shown in Fig. 3 for this monolayer, metal-semiconductor phase transition occurs when biaxial or uniaxial tensile loading along the zigzag is applied. In contrary the metallic electronic character was completely preserved for the sample under uniaxial tensile loading along the armchair. Interestingly the formed band-gap is larger for the uniaxial tensile loading along the zigzag as compared with the biaxial loading. As it is shown, for the biaxial and uniaxial loading along the zigzag borophene hydride monolayer yields direct band-gap semiconductor character in which the valence band maximum (VBM) and conduction band minimum (CBM) lie on the $\Gamma$-point. According to our electronic structure results based on the PBE functional, for the samples under tensile strain of 0.2 for the biaxial straining and uniaxial loading along the zigzag the band-gap is predicted to be 0.12 eV and 0.38 eV, respectively.

Table 1. Summarized energy band-gap (eV) and static dielectric function values of single-layer $B_1H_1$ under different loading conditions.

| Structure | Band-gap | | Re $\varepsilon$ (0) | | |
|---|---|---|---|---|---|
| | PBE | HSE06 | E\|\|x | E\|\|y | E\|\|z |
| Strain=0.0 | metallic | metallic | 1.96 | 1.72 | 1.62 |
| Strain=0.1, biaxial | 0.04 | 0.82 | 2.56 | 2.00 | 1.65 |
| Strain=0.2, biaxial | 0.12 | 1.26 | 3.07 | 2.34 | 1.67 |
| Strain=0.1, uniaxial armchair | metallic | metallic | 2.10 | 1.64 | 1.69 |
| Strain=0.2, uniaxial armchair | metallic | metallic | 2.21 | 1.60 | 1.76 |
| Strain=0.1, uniaxial zigzag | metallic | metallic | 2.11 | 2.01 | 1.70 |
| Strain=0.2, uniaxial zigzag | 0.38 | 1.6 | 2.27 | 2.31 | 1.74 |

Since the PBE functional underestimates the band-gap values, we also computed the EDOS using the HSE06 [62] hybrid functional with a 10×10×1 k-point mesh size. The band-gap values predicted by the PBE and HSE06 methods for strained monolayers are reported in Table 1. Because of the fact that the HSE06 method provides more accurate predictions for the band-gap [63], in Fig. 4 we specifically analyze the evolution of band-gap on the basis of HSE06 results. As it is clear, as compared with the uniaxial loading along the zigzag, the band-gap opening occurs in a more uniform pattern for the biaxial loading. This observation highlight that the biaxial straining can be considered as a more promising route for the engineering of electronic band-gap in borophene hydride. Worthy to remind that for the flat and buckled pristine borophene nanomembranes, mechanical loading could not lead to the band-gap opening [34]. Presenting a small band-gap semiconductor electronic character is a very promising feature for the application of this novel 2D structure in nanoelectronics.

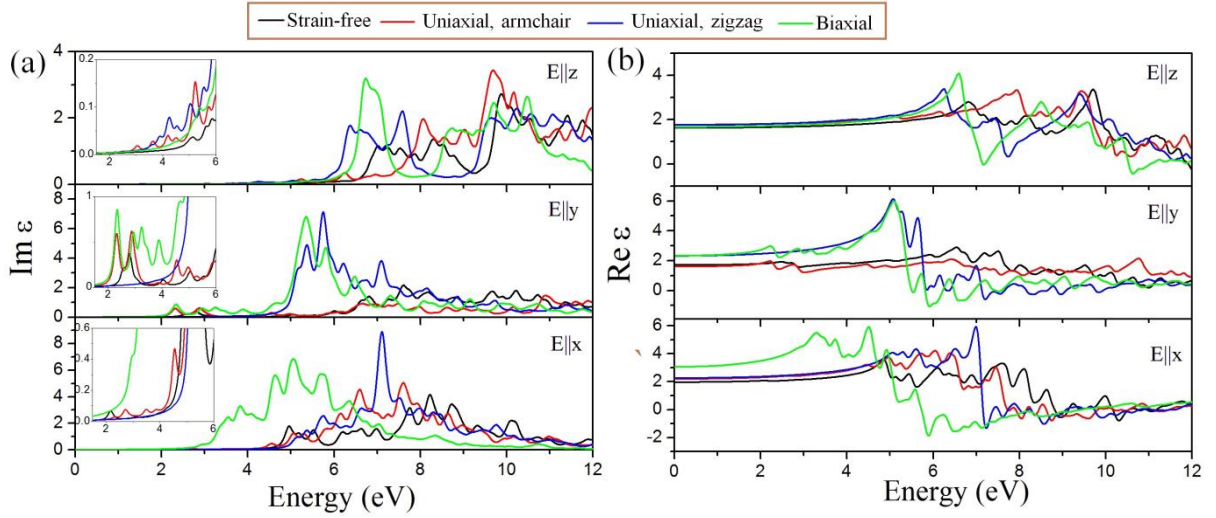

Fig. 5, Imaginary and real parts of the dielectric function of single-layer borophene hydride with no strain and under the strain level of 0.2 for the parallel (E‖x and E‖y) and perpendicular (E‖z) light polarizations, calculated using the PBE plus RPA approach. Insets show amplified regions of $\text{Im}\varepsilon_{\alpha\beta}(\omega)$ in the low frequency regime.

We next analyze the optical response of this novel 2D material. The imaginary and real parts of the dielectric function of strain-free and strained monolayers for the parallel and perpendicular polarized directions ($E//x$, $E//y$ and $E//z$) are illustrated in Fig. 5. In this case for the strained systems, we only considered the structure with large strain of 0.2 as it shows more considerable effects on the electronic response. For strain-free structure, the absorption edge of $\text{Im}\varepsilon_{\alpha\beta}(\omega)$ is at 1.90, 2.15 and 4.60 eV for $E//x$, $E//y$ and $E//z$, respectively, which are in visible range of light for parallel polarization. These results indicate that these 2D systems can absorb visible light. Moreover, there are small peaks in low frequency regime for the parallel polarizations which is completely missing in perpendicular one. The value of the static dielectric constant (the real part of the dielectric constant at zero energy) is 1.96, 1.72 and 1.62 for $E//x$, $E//y$ and $E//z$, respectively. Since our previous electronic calculations indicated that the band-gap of this system can be increased by exerting biaxial or uniaxial loading along the zigzag, we also calculated the Im $\varepsilon$ and Re $\varepsilon$ for the strained structures. In Fig. 5, the anisotropic response for parallel and perpendicular polarized directions to the planes can be easily observed for the all strained structures. It can be seen that for the parallel polarizations, by applying uniaxial loading along the zigzag the adsorption edge of Im $\varepsilon$ in the low-frequency regime shifts to the higher energies, which is known as blue shift. On the other side, a shift to lower energies (red shift) is observable by applying biaxial strain or uniaxial loading along the armchair. In addition, by applying biaxial or uniaxial loading along the zigzag the value of the static dielectric constant in all polarizations increases while for the uniaxial loading along the armchair it increases along E‖x and E‖z and decreases along E‖y. Table 1 also summarizes the value of static dielectric constant for all explored cases.

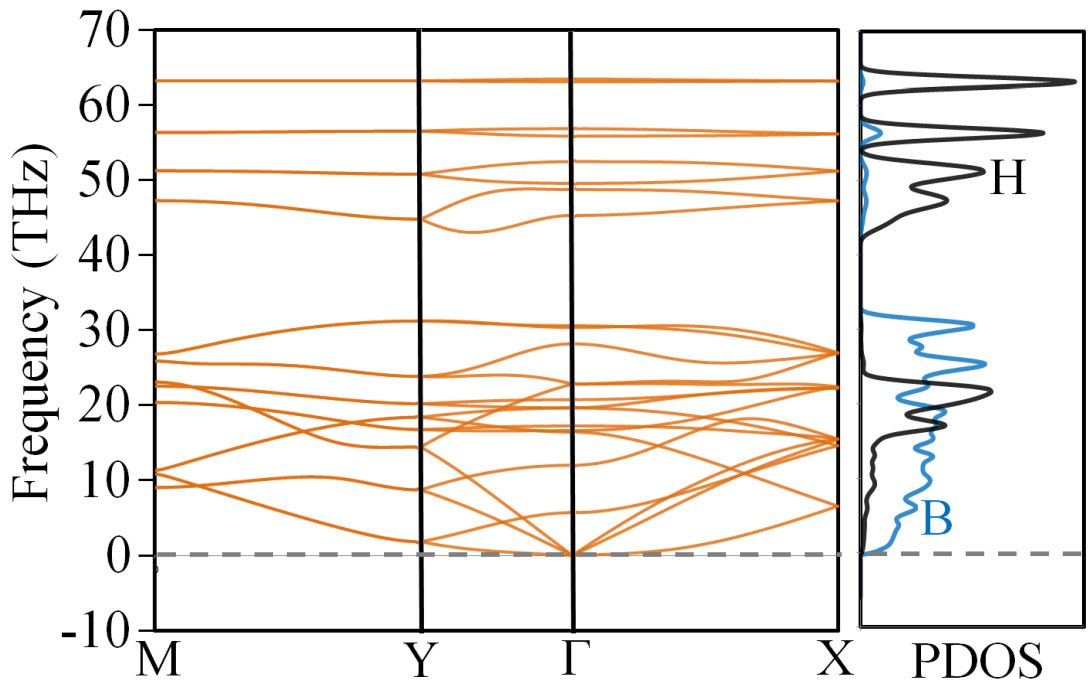

Fig. 6, Phonon dispersion and partial phonon density of states (PDOS) of single-layer borophene hydride.

We finally analyze the lattice (phononic) thermal transport of borophene hydride. Phonon dispersion curve of borophene hydride is presented in Fig. 6, which shows an excellent agreement with the earlier investigation [46]. Around the center of the Brillouin zone (Γ point) we notice three acoustic modes, two of them with linear dispersion and a third one with parabolic dispersion, characteristic of 2D materials. We also notice the gap in the phonon bands, in the interval from approximately 30 to 42 THz, characteristic of the mass difference in the chemical species present, B and H. The vibrational density of states is also presented in the figure, projected on B atoms and H atoms, respectively. In the lower frequency range, i.e. below the phonon gap, the phonon density of states (PDOS) is dominated by B atoms, whereas in the higher frequency range (above the gap) it is dominated by H atoms. This behaviour is due to the mass difference, since hydrogen atoms have lower mass and therefore tend to vibrate with higher frequencies. Our results also indicate no sign of imaginary (negative) frequencies in the phonon dispersion which accordingly confirms the dynamical stability of this attractive 2D material.

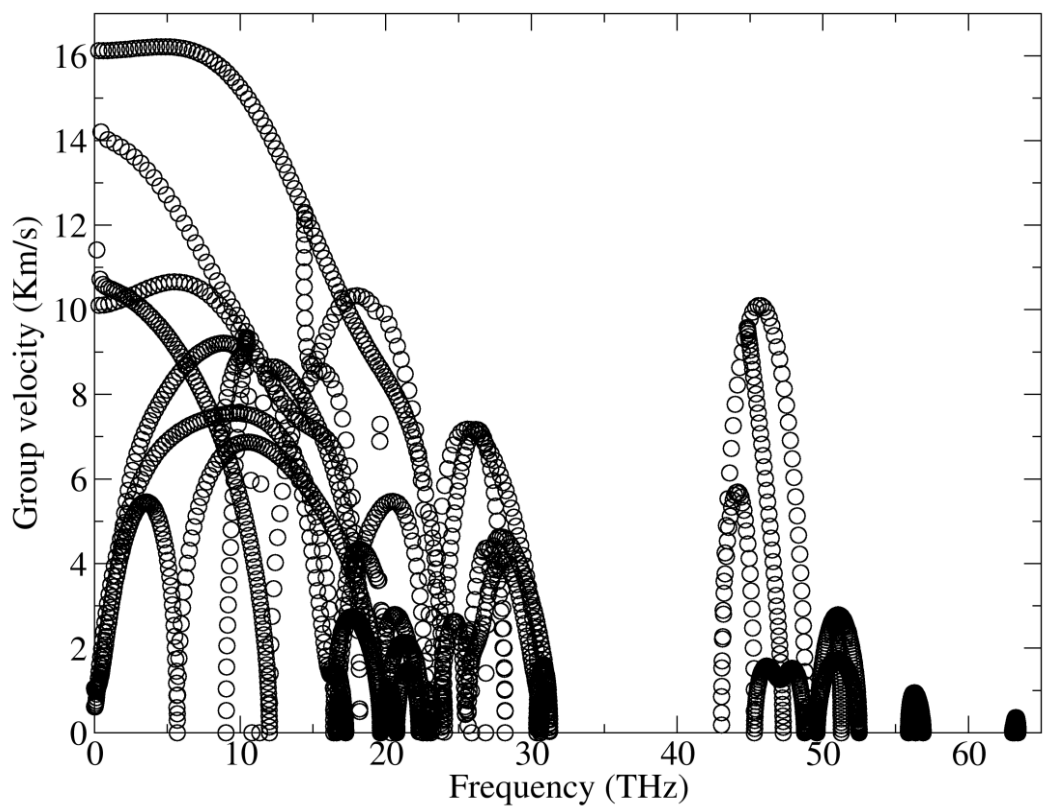

Fig. 7, Predicted phonon group velocity of single-layer borophene hydride.

Next, we consider the phonon group velocities for all vibrational modes as depicted in Fig. 7. In this case, we used a q-point path identical to the one employed in the calculation of phonon dispersion shown in Fig. 6. Such a presentation can be very illustrative to observe the group velocity associated with the acoustic modes. This way inside the region of phonon gap, the group velocity is undefined. Based on our results shown in Fig. 7, the group velocities of the acoustic modes achieve values in excess of $10^4$ m/s, which indicates a probable high thermal conductivity and long effective phonon mean free paths.

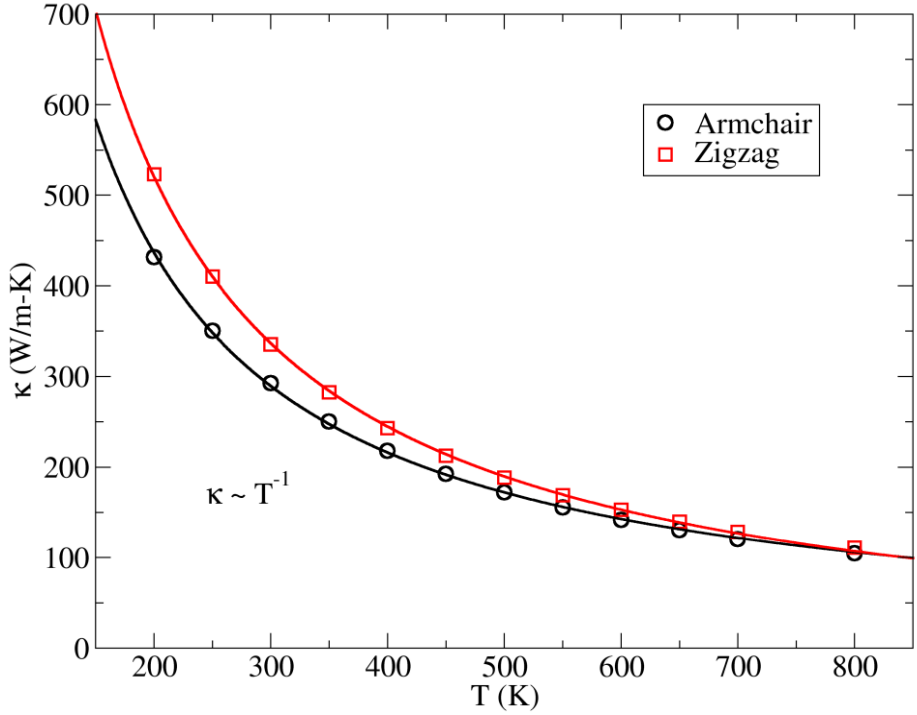

Fig. 8, Temperature dependent lattice thermal conductivity of single-layer borophene hydride for the heat transfer along armchair and zigzag direction. The thickness was assumed to be 4.16 Å which is based on the van der Waals diameter of boron atom.

We calculated the lattice thermal conductivity along two in-plane directions, armchair and zigzag in analogy with graphene, in the temperature interval from 200 K to 800 K and the results are illustrated in Fig. 8. In agreement with the previous study [64], the van der Waals diameter of boron atom (4.16 Å) was taken as the thickness of single-layer borophene. We notice an anisotropy in the conductivity, which is particularly pronounced in the lower temperature regime. Over the temperature range considered the conductivity along the zigzag direction is larger than the one along the armchair direction. Nonetheless, the thermal conductivity along both directions is inversely proportional to temperature, T, as expected when phonon-phonon scattering dominates over phonon-defect, and phonon-boundary scattering [65]. Interestingly the observed anisotropy in the thermal conductivity is inverse to that for the elastic modulus. The origin of anisotropy in the both load transfer and phonon transport can be partially attributed to the B-H bonds, which yield contrasting effects; stiffening effect for the load transfer and scattering effect for the phonon transport. Nevertheless, based on our first-principles results single-layer borophene hydride at room temperature can exhibit high thermal conductivity of 335 W/mK and 293 W/mK along the zigzag and armchair direction, respectively. These values are by around 50 % higher than the earlier prediction for Pmmn borophene structure [66] which is reasonable since the Pmmn borophene shows buckled atomic structure.

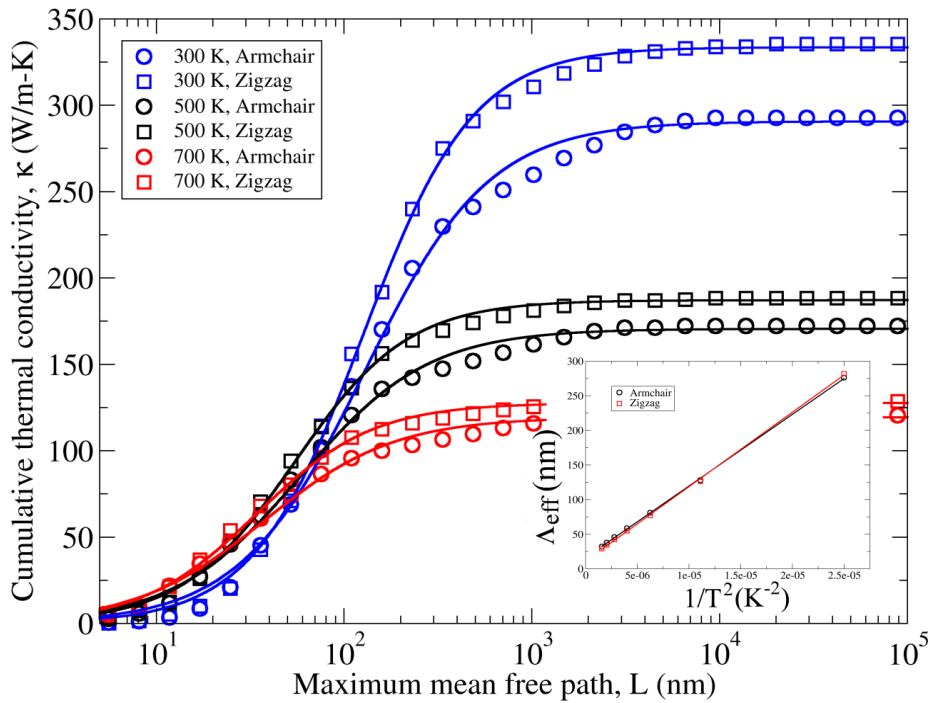

Fig. 9, The accumulative lattice thermal conductivity of borophene hydride along the armchair and zigzag as a function of phonon mean free path at different temperatures of 300 K, 500 K and 700 K. Inset shows the dependence of the effective phonon mean free path with respect to the temperature.

We analyze the dependence of the cumulative lattice thermal conductivity with respect to the maximum phonon mean free path in the samples, which corresponds to the system length along the heat transport direction. From the behaviour of the cumulative conductivity with the system length we can estimate an effective phonon mean free path along each direction in a range of temperatures, using the following expression [67,68]:

$$\kappa(L) = \kappa_\infty / \left[1 + (\Lambda_{eff}/L)^\alpha\right] \quad (3)$$

Here, the $\Lambda_{eff}$ is the effective phonon mean free path, $L$ is the length and $\kappa_\infty$ is the length independent thermal conductivity. In the above equation we introduced an extra parameter, α, temperature-dependent exponent "α", which modifies the influence of the system length on the calculated conductivity for each temperature. In the Fig. 9 we show the cumulative lattice thermal conductivity as a function of maximum phonon mean free path for the both studied directions at 300 K, 500 K and 700 K. The data points are calculated by ShengBTE, while the continuous lines are fits based on the equation above. With the inclusion of the extra exponent "α", we see an agreement of almost four decades between calculation data and fit. Notice that previous works have always considered α=1, but in our case it decreases with temperature from 1.33 to 1.25 for the armchair direction and from 1.50 to 1.40 along the zigzag direction. Nonetheless, it is interesting to notice that α is only weakly dependent on the temperature. Finally, we investigate the dependence of $\Lambda_{eff}$ with temperature. In contrast with the lattice thermal conductivity, there is no clear prediction for the dependence of the effective phonon mean free path with temperature. Adjusting different functional forms to the data points from 200 K to 800 K we found $\Lambda_{eff} \sim T^{-2}$, as shown in the Fig. 9 inset. To the best of our knowledge this is the first time that this kind of behaviour has been observed. Furthermore, the behaviour does not depend on the heat current direction, just as in the case of the temperature dependence of lattice thermal conductivity.

## 4. Conclusion

Most recently, a hydrogenated borophene structure with an empirical formula of $B_1H_1$ was realized. In this work we elaborately explored the mechanical response, thermal conductivity, electronic and optical responses of borophene hydride using the first-principles density functional theory calculations. The mechanical properties of borophene hydride were found to be anisotropic in which the elastic modulus along the armchair and zigzag directions were predicted to be 131 N/m and 99 N/m, respectively. It was also shown that borophene hydride can yield remarkable tensile strengths of 19.9 N/m and 17.8 N/m along the armchair and zigzag directions, respectively. As an interesting finding, it was shown that by applying biaxial tensile strains or uniaxial tensile loading along the zigzag direction the metallic electronic character of borophene hydride can be changed to direct band-gap semiconducting character, which is a highly desirable factor for the application in nanoelectronics. Our results therefore confirm that tensile strains can be considered as a promising route for the engineering of electronics properties of borophene hydride. For the optical properties of this novel 2D material, we found that the absorption edge of the imaginary part of the dielectric function happen in the visible range of light for parallel polarization. Based on our first-principles results, it was predicted that the single-layer borophene hydride at the room temperature can exhibit high thermal conductivities of 335 W/mK and 293 W/mK along the zigzag and armchair direction, respectively. The high thermal conductivity of borophene hydride is very appealing for application in thermal management systems and enhancing the thermal conductivity of polymeric materials. Based on our results the anisotropy in the thermal conductivity is inverse to that observed for the elastic response. It was concluded that the B-H bonds can partially act as the origin of anisotropy in both mechanical and thermal conduction properties, they simultaneously enhance the load bearing of the structure and increase the phonon scattering rate. This investigation provides comprehensive information concerning the mechanical, electronic, optical and thermal conduction properties of borophene hydride and such that it can be useful for the future theoretical and experimental studies.

**Acknowledgment**

B. M. and T. R. greatly acknowledge the financial support by European Research Council for COMBAT project (Grant number 615132). C.V.S. and M.M. gratefully acknowledge their financial support in parts by Natural Sciences and Engineering Council of Canada (NSERC), University of Toronto, Connaught Global Challenge Award, and Hart Professorship. L.F.C.P. acknowledges financial support from CAPES for the project "Physical properties of nanostructured materials" (Grant No. 3195/2014) via its Science Without Borders program and provision of computational resources by the High Performance Computing Center (NPAD) at UFRN.